\documentclass[letterpaper, 10 pt, conference]{ieeeconf}  
\IEEEoverridecommandlockouts                              
\usepackage{amsmath, amssymb, mathtools}  
\usepackage{lipsum}  
\usepackage{gensymb}
\usepackage{wrapfig}
\usepackage{mathrsfs}
\usepackage[caption=false,font=footnotesize]{subfig}
\usepackage{wrapfig}
\usepackage{breqn}
\usepackage{amsfonts}
\usepackage{graphicx}
\usepackage{xcolor}

\title{\LARGE \bf
A state-space model for dynamic functional connectivity}

\author{Sourish Chakravarty$^{1,5,6}$, Zachary D. Threlkeld$^{7}$, Yelena G. Bodien$^{6}$, Brian L. Edlow$^{6}$, Emery N. Brown$^{1-5}$ 
\thanks{This work was partially supported by NIH Award P01-GM118629 (to E.N.B), by funds from Massachusetts General Hospital (to E.N.B.), by funds from the Picower Institute for Learning and Memory (to E.N.B. and S.C.), by funds from the NIH/NINDS (DP2-HD101400, R21-NS109627, RF1-NS115268, K23-NS094538; to B.L.E.), by funds from James S. McDonnell Foundation (to B.L.E.), and by funds from the National Institute on Disability, Independent Living and Rehabilitation Research, Administration for Community Living (90DP0039, Spaulding- Harvard TBI Model System; to Y.G.B.).
}
\thanks{$^{1}$Picower Institute for Learning and Memory, Massachusetts Institute of Technology (MIT), Cambridge, MA}%
\thanks{$^{2}$Department of Brain and Cognitive Sciences, MIT, Cambridge, MA}%
\thanks{$^{3}$Institute of Medical Engineering and Science, MIT, Cambridge, MA}%
\thanks{$^{4}$ Harvard-MIT Division of Health Science and Technology}%
\thanks{$^{5}$ Department of Anesthesia, Critical Care and Pain Medicine, Massachusetts General Hospital (MGH), Boston, MA}%
\thanks{$^{6}$ Center for Neurotechnology and Neurorecovery, Department of Neurology, MGH, Boston, MA}%
\thanks{$^7$ Dept. of Neurology and Neurological Sciences, Stanford University School of Medicine, Stanford, CA}
\thanks{Accepted in 53rd Annual Asilomar Conference on Signals, Systems, and Computers, Pacific Grove, CA. $\copyright$ 2019 IEEE. Personal use of this material is permitted. Permission from IEEE must be obtained for all other uses, in any current or future media, including reprinting/republishing this material for advertising or promotional purposes, creating new collective works, for resale or redistribution to servers or lists, or reuse of any copyrighted component of this work in other works.}
}
\begin{document}
\maketitle

\begin{abstract}
Dynamic functional connectivity (DFC) analysis involves measuring correlated neural activity over time across multiple brain regions. Significant regional correlations among neural signals, such as those obtained from resting-state functional magnetic resonance imaging (fMRI), may represent neural circuits associated with rest. The conventional approach of estimating the correlation dynamics as a sequence of static correlations from sliding time-windows has statistical limitations. To address this issue, we propose a multivariate stochastic volatility model for estimating DFC inspired by recent work in econometrics research. This model assumes a state-space framework where the correlation dynamics of a multivariate normal observation sequence is governed by a positive-definite matrix-variate latent process. Using this statistical model within a sequential Bayesian estimation framework, we use blood oxygenation level dependent activity from multiple brain regions to estimate posterior distributions on the correlation trajectory. We demonstrate the utility of this DFC estimation framework by analyzing its performance on simulated data, and by estimating correlation dynamics in resting state fMRI data from a patient with a disorder of consciousness (DoC). Our work advances the state-of-the-art in DFC analysis and its principled use in DoC biomarker exploration.
 \end{abstract}
\section{Introduction}
Functional connectivity (FC) analysis can be broadly described as an analysis of ``statistical dependencies among remote neuro-physiological events" \cite{friston2011functional}. The FC measures are considered to be representative of long-range neural coordination across brain-regions. Conventionally, they are calculated as pair-wise correlation in neural signals, e.g. blood oxygen level dependent (BOLD) activation in functional magnetic resonance imaging (fMRI) data and or electrical activity in electroencephalography (EEG) data, across multiple brain regions. From a signal processing perspective, FC analysis is based upon estimation of correlation matrices from multivariate time-series data. FC analysis is widely relevant in basic neuroscience \cite{antzoulatos2014increases} as well as translational clinical neuroscience research (\cite{bodien2017functional,Edlow2017Brain}). 

An emerging paradigm in FC research is that of dynamic functional connectivity (DFC) \cite{Tagliazucchi2015Frontiers}. Conventionally, FC analysis is performed on an entire session of multi-site/multi-channel neural recordings to generate a correlation map. A key assumption of such an analysis is that the true correlation remains constant across the entire sequence of a recording session. This static FC (SFC) interpretation can be generalized to DFC by assuming that the true correlation map  changes over time. Therefore, DFC analysis involves estimating a time-series of pairwise correlations from multivariate neural data. The DFC analysis yields information that is complementary to SFC analysis of the same dataset: SFC indicates the degree to which any two brain-regions are synchronous with each other on average, whereas DFC indicates how this degree of synchrony varies over time. Principled DFC analysis may aid translational research aimed at biomarker discovery for neurological disorders such as disorders of consciousness (DoC) \cite{demertzi2019human}.

Conventionally, DFC analysis generates dynamic estimates of correlations by assuming a sliding-window (SW) approach. In this approach, a sequence of instantaneous correlation measures is calculated by repeating the SFC analyses on a sequence of short data windows where the window-size is user-prescribed. A sub-optimal choice of window-size can lead to statistical challenges. For example, if the window size is too long, then the estimated correlation trajectory may be unable to capture faster dynamics. Conversely, if window size is too short then the estimates will be more sensitive to noise in the data. Therefore, it is preferable to use alternative DFC estimation techniques that can yield reliable estimates of correlation dynamics without requiring subjective windows. One such promising alternative has been proposed by Lindquist et al.\cite{lindquist2014evaluating} where the authors estimated correlation dynamics by fitting a dynamic conditional correlation model (DCC) to resting-state fMRI data. The DCC model originated in econometrics literature to estimate covariance trajectories in multivariate financial time-series data \cite{engle2001theoretical}. Motivated by this work, here we propose a novel approach based on state-space models (SSMs) to estimate DFC from neural data while avoiding data-windowing. In particular, we consider a specific class of SSMs known as Multivariate Stochastic Volatility (MVSV) models that also originated in econometrics research \cite{casarin2018bayesian}. Such an MVSV model incorporates a matrix-variate latent state process to describe correlation dynamics in the observed neural signals. DFC statistics may then be derived by estimating the latent states and model parameters from the data. This SSM framework provides more  flexibility in describing complex correlation dynamics \cite{chan2016modeling}, and in estimating instantaneous correlations even under scenarios of missing data and non-uniform sampling time-intervals of observations \cite{shumway1982approach}.


In Sec.~\ref{sec:Model_Estimation}, we present our SSM, which is a modified version of the existing MVSV model \cite{casarin2018bayesian}, for estimating correlation dynamics in multivariate noisy neural data. Specifically, we do not consider certain modeling features of \cite{casarin2018bayesian} such as a first-order vector autoregressive process to track the mean of observations, and additional latent states to track both gradual changes in channel-wise variance and temporally sharp regime transitions in mean and covariance dynamics of the signals. Except for minor differences in notational conventions and a major difference in the proposal distributions in Secs.\ref{sec:mcmc_post_nu} and \ref{sec:mcmc_post_d}, the statistical model and estimation procedure used here are identical to \cite{casarin2018bayesian} conditioned on the assumptions made in this work. In Sec.~\ref{sec:Results}, we numerically verify our model on synthetic data and demonstrate its utility in estimating time-varying correlations from BOLD-fMRI data from a DoC patient. Finally, in Sec.~\ref{sec:Conclusion}, we summarize the current work, and discuss potential next steps. 
\section{Theory}\label{sec:Model_Estimation}
\subsection{State-Space Model (SSM)} \label{sec:SSM}
Consider a sequence of multivariate neural data $\lbrace y_k \rbrace_{k=1}^K$. At the $k$-th instant, the observation vector, $y_k\in \mathbb{R}^{m}$, is assumed to be a realization of a multivariate normal distribution characterized by zero mean and covariance matrix $\Sigma_k$. This is represented as,
\begin{align}
    & y_k \sim \mathcal{N}_m(0_m, \Sigma_k)
    \label{eq:SSM_Obs}
\end{align}
where, $0_m$ denotes a $m\times1$ matrix of zeros. The covariance matrix, $\Sigma_k$, is decomposed as,
\begin{align}
    \Sigma_k = \Lambda_k \Omega_k \Lambda_k 
    \label{eq:Bollerslev}
\end{align}
where, $\Lambda_k^2$ is a diagonal matrix of variances and $\Omega_k$ denotes the current correlation matrix. The stochastic evolution of the sequence, $\lbrace \Omega_k \rbrace_{k=1}^{K}$, is assumed to be governed by a latent state sequence, $\lbrace Q_k \rbrace_{k=1}^{K}$ where $Q_k$ is a real-valued positive-definite random matrix. The dynamics of this latent positive-definite process is described by,
\begin{align}
    & Q_k^{-1} = \frac{1}{\nu} Q_{k-1}^{-d/2}E_k Q_{k-1}^{-d/2}\,\, \text{ such that }
    E_k \sim \mathcal{W}_m(\nu, I_m) \,,
    \label{eq:SSM_latent_Q}
    \\
    & \Omega_k = \widetilde{Q}_k^{-1} Q_k\widetilde{Q}_k^{-1} 
    \label{eq:SSM_latent_Om}
\end{align}
where, $\widetilde{Q}_{k,ij}=Q_{k,ij}^{1/2}\delta_{ij}$, $\delta_{ij}$ denotes the Kronecker delta function, $\nu>m$, and $-1\le d \le 1$. Here, we use $X\sim \mathcal{W}_m(\nu, S)$ to denote a real-valued, positive-definite, $m\times m$ random matrix $X$ distributed according to a Wishart probability density function (pdf) described as, 
\begin{align}
  &   \left( 
    2^{\frac{\nu m}{2}}\Gamma_m(\nu/2) \vert S \vert^{\frac{\nu}{2}} 
    \right)^{-1}  \vert X\vert^{\frac{\nu-m-1}{2}}
    \mathrm{eTr}\left( -\frac{1}{2} S^{-1}X
    \right) \,,
    \label{eq:pdf_Wishart}
    \\
 &   \Gamma_m (\nu/2) = \pi^{m(m-1)/4}\prod\limits_{i=1}^{m}
    \Gamma \left( (\nu - i + 1) /2 \right)\,.
    \label{eq:Wishart}
\end{align}
and characterized by parameters $S$ (a positive-definite matrix) and scalar $\nu$ ($>m$) \cite{gupta2009matrix}. In this work, we use $\vert \cdot \vert$ and $\mathrm{eTr}(\cdot)$ to denote, respectively, determinant and exponential-of-trace operations on a square matrix. Note that Eq.~\eqref{eq:SSM_latent_Q} ensures that each point on the dynamic trajectory of $Q_k$ will be positive-definite, and $Q_k^{-1}\sim \mathcal{W}_m(\nu,  Q_{k-1}^{-d}/\nu)$ \cite{gupta2009matrix}. The matrix scaling operation in Eq.~\eqref{eq:SSM_latent_Om} preserves the unit-diagonal character of the resulting correlation matrices, $\Omega_k$. Further simplifying our theoretical modeling in Eq.~\eqref{eq:SSM_latent_Q}, we do not consider additional multiplicative coefficient matrices that can be used in relative scaling of past information ($Q_{k-1}$) \cite[Eqs.8, 9]{casarin2018bayesian}, and assume $\Lambda_k =I_m$, where $I_m$ denotes a $m\times m $ identity matrix.
\subsection{Likelihood and model parameters}
The total data log-likelihood can be written out as,
\begin{align}
    p(y_{1:K}, Q_{1:K}\vert \Theta) = 
    p(Q_0)\prod\limits_{k=1}^{K} p(y_k\vert Q_k) p(Q_k \vert Q_{k-1}, \Theta)
\end{align}
where, $p(X)$ denotes a pdf described on the appropriate space to which $X$ belongs, and $\Theta=\lbrace \nu, d \rbrace$. We assume $Q_0 = I_m$ and set $p(Q_0)=1$ \cite{casarin2018bayesian}. 
\subsection{Bayesian inference \& Markov Chain Monte Carlo steps}
For a given sequence of observations, $y_{1:K}\equiv \lbrace y_k \rbrace_{k=1}^K$, we employ a Bayesian inference framework to estimate posterior distributions on the DFC statistics, $Q_{1:K}^{-1}\equiv \lbrace Q_k^{-1}\rbrace_{k=1}^K$ and $\Theta$. Using the model (Eqs.~\eqref{eq:SSM_Obs}, \eqref{eq:Bollerslev}, \eqref{eq:SSM_latent_Q},  \eqref{eq:SSM_latent_Om}), the unit variance assumption, Bayes' rule and prior probability distributions $p(\nu)$ and $p(d)$, we can write the following posterior pdf, 
\begin{align}
    p(Q_{1:K}^{-1}, \nu, d\vert y_{1:K}) \propto 
    p( y_{1:K}, Q_{1:K}^{-1} \vert \nu, d) p(\nu) p(d),
    \label{eq:pdf_post_joint}
\end{align}
where, 
\begin{align}
    & 
    p(\nu) =  \frac{\beta_\nu^{\alpha_\nu}}{ \alpha_{\nu}}
    (\nu-m)^{\alpha_\nu -1} \mathrm{exp}( -\beta_{\nu} (\nu-m) )
    \mathbb{I}_{(0,\infty)}(\nu-m),
    \label{eq:priorpdf_nu}
    \\
    & 
    p(d) = (1/2)\mathbb{I}_{[-1,1]}(d),
    \label{eq:priorpdf_d}
\end{align}
where, $(\alpha_\nu, \beta_\nu)$ denote the user-prescribed hyper-parameters for the prior pdf on $(\nu -m)$ distributed according to a Gamma probability distribution denoted as $Gamma(\alpha_\nu, \beta_\nu)$.  $\mathbb{I}_{(\mathcal{X})}(x)$ denotes an indicator function that takes value $=1 \,\forall x\in \mathcal{X}$ and $=0$, otherwise. 

To sample from the joint posterior pdf (Eq.~\eqref{eq:pdf_post_joint}) per \cite{casarin2018bayesian}, we use a Gibb's sampling approach \cite{chib1995understanding} where in the $j$-th iteration we draw the $j$-th samples from the following conditional posterior pdf's,
\begin{align}
    Q_{k,(j)}^{-1} \sim
    p( Q_{k-1}^{-1} \vert Q_{k-1,(j)}^{-1}, Q_{k+1,(j-1)}^{-1}, \nu_{(j-1)}, d_{(j-1)}, y_{1:K}) 
    \label{eq:postpdf_Q}
\end{align}
\begin{align}
    & Q_{K,(j)}^{-1} \sim
    p( Q_{K}^{-1} \vert Q_{K-1,(j)}^{-1}, \nu_{(j-1)}, d_{(j-1)}, y_{1:K}) 
    \label{eq:postpdf_Q_K}
        \\
    & \nu_{(j)} \sim
    p(\nu \vert Q_{1:K, (j)}^{-1}, d_{(j-1)}, y_{1:K}) =
    p(\nu \vert Q_{1:K, (j)}^{-1}, d_{(j-1)})
    \label{eq:postpdf_nu}
        \\
    & d_{(j)} \sim
    p(d \vert Q_{1:K, (j)}^{-1}, \nu_{(j)}, y_{1:K}) =   
    p(d \vert Q_{1:K, (j)}^{-1}, \nu_{(j)}) 
    \label{eq:postpdf_d}
\end{align}
where, the equalities in Eqs.~\eqref{eq:postpdf_nu} and \eqref{eq:postpdf_d} are due to the conditional independence assumption in the proposed SSM. To sample $x^{(j)}\sim p(x) $ (where, $p(x)$ may denote any of the $K+2$ conditional pdf's in Eqs.~\eqref{eq:postpdf_Q}-\eqref{eq:postpdf_d}), given the last sample $x^{(j-1)}$, we perform the following general steps of MCMC algorithm \cite{chib1995understanding}: we draw a candidate sample $x^{\star}$ from a relatively easy-to-sample \textit{proposal} distribution whose pdf, conditioned on $x^{(j-1)}$, is denoted by $q(x^{\star}\vert x^{(j-1)})$ and accept it as a new sample $x^{(j)}$ with an acceptance probability given by,
\begin{align}
    \min \left( 
    1, \frac{
    g(x^\star) q(x^{(j-1)}\vert x^\star)
    }{
    g(x^{(j-1)}) q(x^{\star}\vert x^{(j-1)})
    }
    \right)
    \label{eq:MCMC_acceptance}
\end{align}
where, $p(x) \propto g(x)$. If we are unable to accept $x^\star$, then we set $x^{(j)}= x^{(j-1)}$. In the following sections, Secs.~\ref{sec:mcmc_post_Qk} - \ref{sec:mcmc_post_d}, we provide the details of our choice of $g(\cdot)$ and $q(\cdot \vert \cdot)$ that will be used in Eq.~\ref{eq:MCMC_acceptance} for each of the conditional posterior pdfs, Eqs.~\eqref{eq:postpdf_Q}-\eqref{eq:postpdf_d}. Note that with the expressions available to calculate $g(x^\star)$ and $q(x^\star\vert x^{(j-1)})$, we can also calculate $g(x^{(j-1)})$ and $q(x^{(j-1)}\vert x^\star )$ by exchanging $x^{(j-1)}$ and $x^\star$ in the arguments. 
\subsection{Sampling $Q_k^{-1}$ from conditional posterior pdf}\label{sec:mcmc_post_Qk} 
The conditional posterior in Eq.~\eqref{eq:postpdf_Q} can be evaluated up to an unknown constant factor as,
\begin{align}
    &  p(y_k \vert Q_k^{-1}, \Theta) 
    p(Q_{k+1}^{-1}\vert Q_{k}^{-1}, \Theta)
    p(Q_{k}^{-1}\vert Q_{k-1}^{-1}, \Theta) 
    \propto 
    g( Q_k^{-1} ) 
    \label{eq:pdf_Qkm1_post_1}
    \\
    & = \vert Q_k^{-1} \vert ^{(\nu + 1 -m-1)/2}
    \mathrm{eTr}\left(  -\frac{\nu}{2}S_{k}^{-1}Q_k^{-1}\right)
    \vert Q_k^{-1} \vert^{(-d\nu)/2}  \cdot 
    \nonumber
    \\
    & 
    \vert \widetilde{Q}_k \vert 
    \mathrm{eTr}\left( 
    -\frac{1}{2}( 
    (\widetilde{Q}_k \Lambda_k^{-1} y_k y_k^T \Lambda_k^{-1} \widetilde{Q}_k)Q_k^{-1}
    + \nu S_{k+1}^{-1}Q_{k+1}^{-1}
    ) \right)
    \label{eq:pdf_Qkm1_post_2}
    \\
    & =
    \vert Q_k^{-1} \vert ^{(\nu + 1 -m-1)/2} \cdot
    \nonumber
    \\
    &
    \mathrm{eTr}\left(  -\frac{1}{2}(
    \nu S_{k}^{-1} + 
    \widetilde{Q}\Lambda_k^{-1} y_k y_k^{T} \Lambda_k^{-1}\widetilde{Q} 
    ) Q_k^{-1}
    \right)
    \vert Q_k^{-1} \vert^{(-d\nu)/2}  \cdot 
    \nonumber
    \\
    & 
    \vert \widetilde{Q}_k \vert \frac{
    \mathrm{eTr}\left( 
    -\frac{1}{2}( 
    (\widetilde{Q}_k \Lambda_k^{-1} y_k y_k^T \Lambda_k^{-1} \widetilde{Q}_k)Q_k^{-1}
    + \nu S_{k+1}^{-1}Q_{k+1}^{-1}
    ) \right)
    }{
    \mathrm{eTr}\left( -\frac{1}{2}( 
    \widetilde{Q} \Lambda_k^{-1} y_k y_k^{T}
    \Lambda_k^{-1} \widetilde{Q} 
    )Q_k^{-1}
    \right)
    }
\label{eq:pdf_Qkm1_post_3}
\end{align}
where, $S_{k}\equiv Q_{k-1}^{-d}$. Note that Eq.~\eqref{eq:pdf_Qkm1_post_3} is obtained by multiplying and dividing Eq.~\eqref{eq:pdf_Qkm1_post_2} by $\mathrm{eTr}(-(1/2)  ( 
    \widetilde{Q} \Lambda_k^{-1} y_k y_k^{T}
    \Lambda_k^{-1} \widetilde{Q} 
    )Q_k^{-1})$ where $\widetilde{Q}= (\widetilde{Q}_{k-1}+ \widetilde{Q}_{k+1})/2$.
Per \cite{casarin2018bayesian}, to sample a candidate realization, $Q_{k,\star}^{-1}$ we use as proposal pdf, $q(Q_{k,\star}^{-1}\vert Q_{k,(j-1)}^{-1})$, the pdf of $\mathcal{W}_m(\nu + 1, (\nu S_k^{-1} + \widetilde{Q}\Lambda_k^{-1} y_k y_k^T \Lambda_k^{-1}\widetilde{Q})^{-1})$. 
\subsection{Sampling $Q_K^{-1}$ from conditional posterior pdf}\label{sec:mcmc_post_QK}
The conditional posterior in Eq.~\eqref{eq:postpdf_Q_K} can be expressed upto an unknown constant as,
\begin{align}
    &g(Q_{K}^{-1})  = \vert S_K^{-1}\vert^{\nu/2} 
    \vert Q_K^{-1} \vert^{(\nu+1-m-1)/2} 
    \mathrm{eTr}\left( -\frac{\nu}{2} S_K^{-1}Q_K^{-1}\right)\cdot 
    \nonumber
    \\
    &
    \vert \widetilde{Q}_K \vert
    \mathrm{eTr}\left( 
    -\frac{1}{2}
    (\widetilde{Q}_K \Lambda_K^{-1} y_K y_K^T 
    \Lambda_K^{-1} \widetilde{Q}_K
    \right)Q_K^{-1}
    \label{eq:pdf_QK_post}
\end{align}
. Per \cite{casarin2018bayesian}, to sample any candidate $Q_{K,\star}$ we use as proposal pdf, $q(Q_{K,\star} \vert Q_{K,(j-1)})$, the pdf corresponding to $\mathcal{W}_m(\nu + 1, S_K/\nu)$. 
\subsection{Sampling $\nu$ from conditional posterior pdf} \label{sec:mcmc_post_nu}
The conditional posterior distribution in Eq.~\eqref{eq:postpdf_nu} can be written out up to a constant multiplying factor as $g(\nu)$, such that 
\begin{align}
\ln g(\nu) 
    &= 
    \mathbb{I}_{(m, \infty)}( \ln (\nu - m)^{\alpha_\nu -1}
    - K \ln \Gamma_m(\nu/2)
    \nonumber
    \\
    &
    + (m\nu K/2) \ln \nu 
    - (\nu/2) ( 
    2\beta_\nu + mK\ln 2 
    \nonumber
    \\
    &
    + 
    \sum\limits_{k=1}^K (
    \ln \vert S_k \vert - \ln \vert Q_k^{-1}\vert + \mathrm{Tr}(S_k^{-1}Q_k^{-1})  
    )
    )
    )
    \label{eq:logpdf_post_nu}
\end{align}
To generate a candidate sample $\nu^\star$ we use as proposal pdf, $q(\nu^\star \vert \nu^{(j-1)})$, the pdf corresponding to a shifted Gamma distribution denoted by $Gamma(\alpha_p, \beta_p)$. The  characteristic parameters $(\alpha_p, \beta_p)$ are functions of a prescribed mode $\nu_M=\nu^{(j-1)}$ and user-prescribed constant variance $\nu_{var}$, such that 
\begin{align}
    & 
    \beta_{p}(\nu_M, \nu_{var}) = \frac{
    ( (\nu_M - m) + ( (\nu_M - m)^2 + 4\nu_{var} )^{1/2}
    }{
    2\nu_{var}
    \label{eq:proppdf_param_nu_1}
    }
    \\
    & 
    \alpha_{p}(\nu_M, \nu_{var}) = 1 + (\nu_M - m) \beta_{p}(\nu_M)
    \label{eq:proppdf_param_nu_2}
\end{align}
Here, the proposal distribution, $q(\nu^\star \vert \nu^{(j-1)})$, is designed such that the mode of the distribution is at last sample $\nu^{(j-1)}$, and the bounds on $\nu$ are respected. To calculate $q(\nu^{(j-1)} \vert \nu^\star )$, we use a different shifted Gamma pdf whose parameters are calculated using Eqs.~\eqref{eq:proppdf_param_nu_1} and \eqref{eq:proppdf_param_nu_2} with $\nu_M=\nu^{\star}$. 
\subsection{Sampling $d$ from conditional posterior pdf} \label{sec:mcmc_post_d}
The posterior pdf in Eq.~\eqref{eq:postpdf_d} can be written out upto an unknown constant factor as,
\begin{align}
  g(d) &=   \mathbb{I}_{[-1,1]}\exp( 
    \sum\limits_{k=1}^K
    (-d\nu/2) \ln \vert Q_{k-1}^{-1} \vert
    + 
    \nonumber
    \\
    & (-\nu/2) \mathrm{Tr}( Q_{k-1}^{d}Q_{k}^{-1})
    )
    \label{eq:pdf_d_post}
\end{align}
To sample a candidate $d^\star$ we use a proposal density $q(d^\star \vert d^{(j-1)})$ which corresponds to $(1/2) \times$ the pdf of $Beta(a_p,b_p) $. For a given mean, $d_{A}=d^{(j-1)}$, constraint $b_p = 1/a_p$, and a user-prescribed constant threshold parameter $a_f$ such that $1/a_f \le a_p \le a_f$ where $a_f>1$, the characteristic parameter is given by,
\begin{align}
    &a_p(d_{A}, a_f) =
    \nonumber
    \\
    & \max\left(  \min\left(  \left( \frac{ (1+d_A)/2 }{1 - (1+d_A)/2}\right)^{1/2} , a_f \right)  , 1/a_f \right)
    \label{eq:postpdf_d_ap}
\end{align}
The proposal distribution is designed such that its mean corresponds to the previous sample, while also having more probability mass around this mean and respecting the bounds on $d$. To determine $q(d^{(j-1)} \vert d^\star)$, we use $Beta( a_p(d^\star, a_f),1/a_p(d^\star, a_f))$.
\section{Results}\label{sec:Results}
\subsection{Initialization of MCMC simulation}
In all the DFC analyses performed here on synthetic and fMRI datasets, we consider bivariate observation sequences ($m=2$), $K=150$ and maximum MCMC iterations of $N_{MCMC}= 10000$. For the very first MCMC iteration, we initialize $Q_{k,(j=0)}^{-1}={I}_m$, $\forall k = 1,\cdots, K$. The hyper-parameters for the prior on $\nu$, we choose $\alpha_\nu = m + 2$, and $\beta_\nu = 1$, and initialize $\nu_{(j=0)} = m + (\alpha_\nu - 1)/ \beta_\nu$. Also, we set $d_{(j=0)}=0.5$, $\nu_{var}=0.1$ and $a_f = 5$. With regard to post-processing the MCMC generated samples, we set the burn-in iterations for latent states and model parameters as 1000 and 4000, respectively. Also, we set the thinning interval for latent state and model parameters as 100 and 200, respectively. The numbers for burn-in and thinning interval are chosen based on visual inspection of the convergence in the respective Markov chains and the auto-correlation (after ignoring the samples from the burn-in iterations) in each MCMC chain. 
\subsection{DFC analyses of simulated data}\label{sec:Results_simudata}
\begin{figure}
    \centering
    \includegraphics[width=0.45\textwidth]{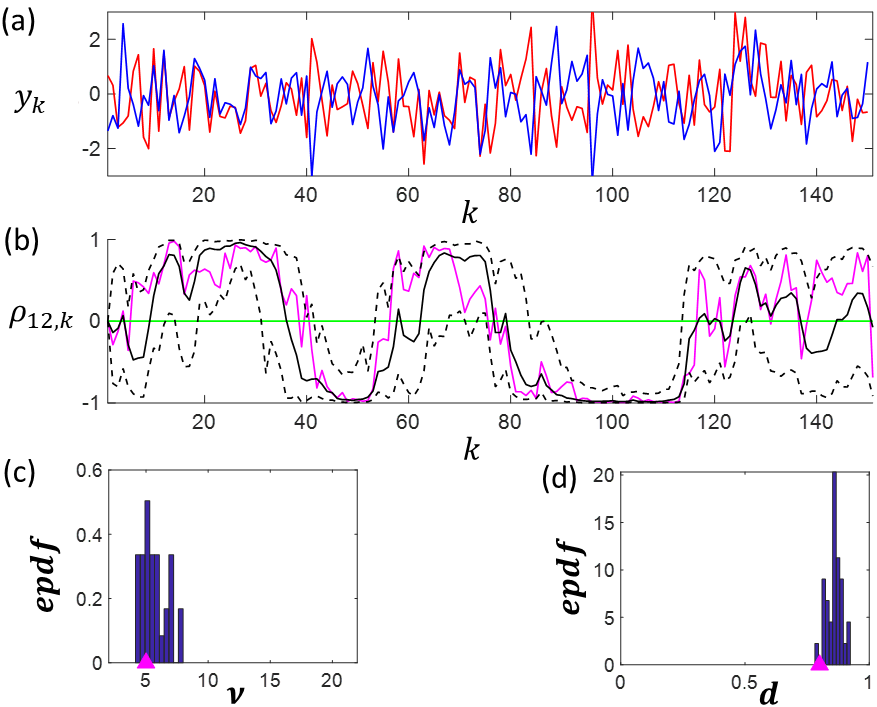}
    \caption{MVSV-based estimation on simulated data. (a) Simulated observations, (b) True correlation (magenta) and ordered statistics on estimated posterior samples (2.5-th, 97.5-th percentiles indicated as dashed black line, while median indicated as solid black line), green line indicates MCMC initialization state. (c) and (d) present estimated empirical pdf's ($epdf$'s) and true value (magenta triangle) on $\nu$ and $d$, respectively.}
    \label{fig:mcmc_simudata}
\end{figure}

Using the generative model (Eqs.\eqref{eq:SSM_Obs}, \eqref{eq:SSM_latent_Q}, \eqref{eq:SSM_latent_Om}, \eqref{eq:Bollerslev}) and setting $\nu = 5$, $d=0.8$, $\Lambda_k = I_2\,\forall \, k = 1,\cdots, K$ we generate a single realization of bivariate observation $y_{1:K}$ (Fig.~\ref{fig:mcmc_simudata}(a)). On this data, we run our MVSV model-based estimation procedure that is agnostic to the ground-truth model parameters. The ordered statistics from samples estimated from the posterior distribution of correlation (Fig.~\ref{fig:mcmc_simudata}(b)), and the empirical histograms on $\nu$ and $d$ (Fig.~\ref{fig:mcmc_simudata}(c,d)) are reported. Note, in Fig.~\ref{fig:mcmc_simudata}(b), the 2.5-th and 97.5-th percentiles include not only the true value, but also track the changes in true correlation trajectory. Comparing with the raw data, Fig.~\ref{fig:mcmc_simudata}(a), we observe that where the estimated 2.5th percentile is above 0, the two time-series appear to be in synchrony (high positive correlation) with each other. On the other hand, when the estimated 97.5th percentile is below 0, the two signals are synchronous, but in the opposite sense (high negative correlation). Therefore, the posterior estimates generated by our DFC analyses enable us to infer whether the correlation is significant at any time instant. In Figs.~\ref{fig:mcmc_simudata}(c) and \ref{fig:mcmc_simudata}(d), the posterior samples of the model parameters are indeed in the near-neighborhood of the ground truth. Also, note that the posterior estimates on $\nu$ and $d$ use only $K=150$ data-points. With longer sessions, these estimates can be expected to be more accurate.  
\subsection{DFC analyses of BOLD fMRI data from a patient}\label{sec:Results_patientdata}
We demonstrate the utility of the DFC analyses pipeline, numerically demonstrated in the previous section, on 2 recording sessions of BOLD-fMRI data. These sessions were collected from a single DoC patient who underwent a resting-state fMRI scan (time to repeat = $2.4$ seconds) while in post traumatic confusional state (PTCS; pre-recovery), and 6 months later after full recovery of consciousness (post-recovery). Details of the clinical trial, recording and data pre-processing procedures can be found in \cite{threlkeld2018functional} \footnote{Denoising, artifact rejection and roi-based averaging steps are performed within CONN Matlab toolbox \cite{whitfield2012conn}}. For the purpose of this proof-of-concept study, we use the denoised and artifact-rejected signal that is averaged within 2 regions-of-interest (ROI's) located in the medial pre-frontal cortex (MPFC) and in the posterior cingulate cortex (PCC). The MPFC and PCC are nodes of the default mode network, and in a healthy subject, these ROI's are known to have positively correlated BOLD activity \cite{bodien2017functional}. The time-series dataset from each of these 2 ROI's is standardized by shifting the mean to zero followed by dividing by the standard deviation, where both the mean and standard deviations are calculated across the entire session. The resulting standardized bivariate sequence constitutes the observations, $y_{1:K}$, to be analyzed.

\begin{figure}
    \centering
    \includegraphics[width=0.45\textwidth]{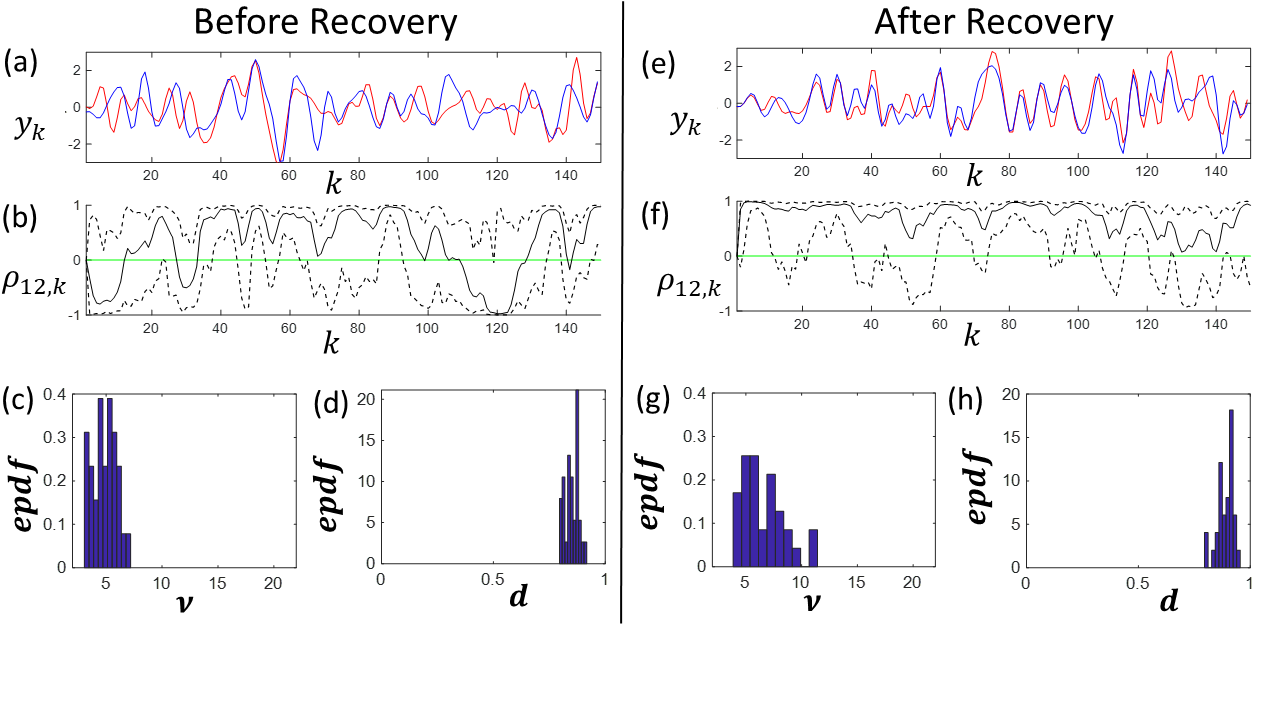}
    \caption{DFC analyses of BOLD fMRI data from a PTCS patient, before and after recovery. (a, e) illustrate the standardized bivariate data from MPFC and PCC, (b, f) present the ordered statistics (2.5-th, and 97.5-th percentiles as dashed black lines, median as solid black line) calculated from the estimated samples of instantaneous correlation. Figure panels (c, g) and (d, h) present empirical pdf's ($epdf$'s) on $\nu$ and $d$, respectively.}
    \label{fig:mcmc_fmri_corr}
\end{figure}

For each session, using our DFC analysis framework we estimate correlation trajectories (Fig.~\ref{fig:mcmc_fmri_corr}) from the entire data sequence without application of a SW approach. The estimated correlation trajectory demonstrates how the degree of synchrony between the BOLD signals from MPFC and PCC gradually evolve over time. The inference of such smooth correlation trajectories is enabled by the MVSV model which incorporates the influence of neighboring time points on the current latent correlation state (Eq.~\eqref{eq:SSM_latent_Q}). Furthermore, this DFC analysis provides session-specific estimates of $\nu$ and $d$, which may be regarded as reduced-order descriptors of the correlation dynamics. For each of these sessions, we note that there are brief epochs where the 95\% confidence interval trajectory of the estimated correlations does not include $0$. Therefore, we infer that for these 2 sessions of fMRI data, the correlation is indeed time-varying. From Fig.~\ref{fig:mcmc_fmri_corr}, we also observe that the correlation trajectory prior to recovery includes both positive and negative correlation values, but the trajectory after recovery remains mostly above zero. Thus, the MVSV model and the corresponding DFC statistics (latent states and model parameters), together provide a principled lower-dimensional objective vocabulary to describe the complex correlation dynamics in multivariate BOLD signals. In the absence of ground truth, the generative nature of the MVSV model allowed us to analyze the performance of the inference framework against synthetically generated ground truth (see Sec.~\ref{sec:Results_simudata}). 
\section{Conclusion}\label{sec:Conclusion}
In summary, we modified an MVSV model from the econometrics literature \cite{casarin2018bayesian} to develop a novel DFC analysis approach for multivariate neural data. We used this approach to perform exploratory DFC analyses of bivariate resting-state BOLD fMRI data from a patient with DoC. We introduced principled modifications to the proposal densities within the MCMC simulations for estimating two key DFC model parameters. We numerically verified the estimation framework on synthetic data. We then deployed the numerically verified DFC analysis pipeline on two fMRI recordings from the patient, before and after recovery of consciousness. Our analysis of resting-state BOLD activity in this DoC patient indicates that the correlation between MPFC and PCC varied within the duration of the recording and correlation dynamics showed marked change when the patient recovered.

Our MVSV model-based DFC analysis pipeline does not require sliding windows, and it can estimate gradually evolving correlation trajectories as well as reduced-order statistical descriptors of correlation dynamics. Furthermore, the estimation framework results in probability distributions, learnt from both prior knowledge and available data, that quantify the uncertainty in both the estimated time-varying correlations and the model parameters. Such probability distributions can be used to compare correlations across multiple time-points in the same session, and to infer dynamic connectivity graphs. A next step will be to extend our bivariate treatment to higher dimensions. Furthermore, this work suggests that exploring SSMs with matrix-variate latent processes for principled DFC analysis of neural data may aid DFC-based bio-marker discovery. 

\bibliographystyle{alpha}
\bibliography{ref}
\end{document}